%% file: 00_archive_QFT_tangentbundle.tex
\documentclass[11pt,aps,amsmath,amssymb]{revtex4}

\usepackage{graphics} 
\usepackage{bm} 
\usepackage{enumerate}
\usepackage{color}

\usepackage[pdftex]{graphicx}

\newcommand{\mymarginpar}[1]{%
  \vadjust{\smash{\llap{\parbox[t]{\marginparwidth}{\; #1}\kern\marginparsep}}}}

\begin{document}
\title{Quantum particles in general spacetimes --- a tangent bundle formalism}
\author{Mattias N.\,R. Wohlfarth}
\email{mnrw2@cantab.net}
\affiliation{{}Universit\"at Hamburg, II. Institut f\"ur Theoretische Physik,\\und 
Goethe-Schule Flensburg, Bismarckstra\ss e 41, 24943 Flensburg, Germany}
\begin{abstract}
Using tangent bundle geometry we construct an equivalent reformulation of classical field theory on flat spacetimes which simultaneously encodes the perspectives of multiple observers. Its generalization to curved spacetimes realizes a new type of non-minimal coupling of the fields, and is shown to admit a canonical quantization procedure. For the resulting quantum theory we demonstrate the emergence of a particle interpretation, fully consistent with general relativistic geometry. The path-dependency of parallel transport forces each observer to carry his own quantum state; we find that the communication of the corresponding quantum information may generate extra particles on curved spacetimes. A speculative link between quantum information and spacetime curvature is discussed which might lead to novel explanations for quantum decoherence and vanishing interference in double-slit or interaction-free measurement scenarios, in the mere presence of additional observers.
\end{abstract}
\maketitle

\tableofcontents

\section{Introduction}
Quantum field theory on Minkowski spacetime admits a clear particle interpretation. On the level of classical field theory, every field satisfies a suitable wave equation that may be solved in terms of a Fourier expansion.
The Fourier modes of the field are eigenmodes of energy and momentum with respect to one timelike and three spacelike Killing vector fields defined in a global Cartesian coordinate system. In the standard quantization scheme the coefficients of the Fourier expansion are then promoted to creation and annihilation operators for particle states, see e.g.~\cite{QFTbook,Kakubook}. On the quantum level, these states transform as representations of the Poincare symmetry group of flat spacetime, an idea going back to Wigner \cite{Wigner:1939cj}. This group encodes both the translational symmetry that provides momentum and mass, and a global Lorentz symmetry that provides us with a well-defined notion of spin for massive fields or helicity for massless fields.

However, this particle interpretation suffers an almost complete break-down when quantum field theory is formulated on curved spacetimes, assuming the standard minimal coupling scheme of fields to the spacetime metric. Now a preferred global coordinate system no longer exists. Even using normal coordinates on a local neighbourhood, the mode solutions of the generalized wave equation do not take a simple form, as for the Fourier expansion on flat spacetime. Killing symmetries generally do not exist, so that a particle interpretation of the mode solutions in terms of four-momenta is heavily obstructed. (The existence of a timelike Killing vector field, or the choice of a timelike observer's worldline tangents, of which the mode solutions are eigenmodes may partially restore the particle energy concept by selecting positive and negative frequencies, but would not be sufficient to restore the momentum interpretation.) 
Different observers may choose different complete sets of mode solutions of the wave equation; though these are related by a Bogolubov transformation, this leads to ambiguities in particle number observations, even on simple spacetimes.
Since there is no global symmetry group for a generic spacetime, also the concept of quantum particle states forming representations is lost. Nevertheless, the effects derived from this framework of ideas, see e.g.~\cite{BirrellDavies},  include famous results like the Davies-Unruh effect~\cite{Davies:1974th,Unruh:1976db} or the Hawking radiation of black holes~\cite{Hawking:1974sw}. The issues arising in the generalization of quantum field theory from flat to curved spacetimes are, of course, well-known, and have led to the rigorous mathematical development of algebraic quantum field theory, where it is accepted that certain features of Minkowski spacetime quantum field theory, like particles, cannot be defined on curved spacetime. Two recent discussions can be found in~\cite{Hollands:2014eia,Fredenhagen:2014lda}.

We will now change our point of view completely. Let us assume that it were possible for any observer to define consistent particle momentum states for quantum fields on curved spacetimes. Then an observer might measure a particle of a certain momentum at some spacetime point; the result of the measurement could be stored by using a gyroscopic device. By general relativistic reasoning the gyroscope axis is then parallely propagated through spacetime along the observer's worldline; see~\cite{Wohlfarth:2012ez} for an extended discussion. A later measurement of the quantum particle state by the same observer should still correspond to the gyroscope reading; this could be expected since the quantum field is spread over the spacetime region where the observer moves, and the particle momentum state is a nonlocal concept. If this expectation holds true, then the particle momentum would not be influenced by any type of motion of the observer, including arbitrary acceleration and rotation. Since vectors at different spacetime points are identified by parallel transport, the particle state would register on the gyroscopic measuring device in the laboratory as totally unchanged. A theory realizing this picture would certainly challenge, and potentially contradict, particle creation results like the above mentioned Davies-Unruh effect or Hawking radiation. 

In this paper we will demonstrate that it is indeed possible to generalize quantum field theory from flat to curved spacetime backgrounds in a way that leads to a consistent particle picture along the lines sketched above. We will achieve this by first, formulating classical field theory on general spacetimes by using techniques from tangent bundle geometry, and second, by quantizing the resulting theory. 

The outline of this paper is as follows. In section~\ref{sec_formulation} we will construct a new formalism for classical field theory on general curved spacetimes, making use of tangent bundle geometry. We review useful conventions and some basic geometry in appendix~\ref{app_basic}. While the new formalism and the standard formalism appear equivalent for flat spacetime, we will find a new non-equivalent generalization of field theory to curved spacetimes. The framework admits a canonical quantization procedure that is performed in section~\ref{sec_quantization}. There, as another key result of this paper, we will also discuss how each observer is enabled to identify the vacuum state and particle momentum states along their worldline, and thus arrive at a quantum particle picture consistent with general relativistic geometry. In section~\ref{sec_infoenergy} we will argue that such a particle picture would result in a direct link between the information about quantum states and energy. Appendix~\ref{app_proof} contains a proof of our particle observation theorem. The paper concludes in section~\ref{sec_discuss} with a summary of our results and an outlook on further possible consequences arising from our new construction.

\section{Tangent bundle formalism for classical field theory}\label{sec_formulation}
This section contains our reformulation of classical field theory in the arena of the tangent bundle of spacetime. We will see that the lift of the field equations to the bundle puts strong emphasis on the observers' perspectives on the physical world. Throughout this paper we will restrict our discussion to the simple case of a scalar field.
After a brief motivation in section~\ref{sec_tmmotivation}  and a look at the flat spacetime case in~\ref{sec_flatmotivation} we will generalize our tangent bundle formulation to curved spacetimes in~\ref{sec_genfieldeq} and~\ref{sec_sfcondition}. Finally, in section~\ref{sec_comparison}, we will wind up with a brief comparison to standard quantum field theory. Some required tangent bundle geometry is reviewed in appendix~\ref{app_basic}.

\subsection{Motivating the case}\label{sec_tmmotivation}
Why tangent bundle geometry enters our reformulation of quantum field theory at all, may be motivated by the following observations. 

The Fourier modes $e^{ip_ax^a}$ of some field $\phi(x)$ in a Cartesian coordinate system on flat spacetime are ill-defined from a geometrical point of view. Indeed, the exponent does not transform as a scalar under diffeomorphisms, since the four-momentum covector is unsuitably contracted with the coordinate functions. Usually this is solved by requiring a much weaker covariance under Lorentz transformations. But this argument cannot be upheld if the flat spacetime theory is to be generalized to curved spacetimes where it is natural to implement general covariance right from the beginning. A solution for this problem would be to consider the quantum field to be a function~$\phi_x(u)$ on some tangent space $T_xM$ with local vector coordinates $(u^a)$. If the equation of motion were written in terms of $u$-derivatives, the associated Fourier modes $e^{ip_au^a}$ would become well-defined scalar quantities. We shall see that the point $x$ then can be interpreted as the spacetime position of the observer of the field $\phi_x(u)$, while $u$ becomes the direction of the line of sight to the point where the field is observed. With this interpretation in mind one realizes also that a Poincare symmetry group acts on the local tangent spaces, since they carry a flat geometry. This makes it plausible why field equations formulated in this arena may allow for a particle interpretation.

Of course, there are many different observers who may measure the same field, momentarily sitting at different points $x$ in spacetime. In fact, already a single observer moves through spacetime along his worldline and will observe the field from his own proper-time dependent location $x(\tau)$ in various directions~$u$. The various field descriptions by functions $\phi_{x(\tau)}:T_{x(\tau)}M\rightarrow \mathbb{R}$ on single tangent spaces at fixed spacetime positions $x(\tau)$ will have to be glued together suitably to form a field function $\Phi:TM\rightarrow \mathbb{R}$ on the tangent bundle of spacetime.

\subsection{Flat spacetime scenario}\label{sec_flatmotivation}
Before considering general spacetimes, we will first discuss the new ingredients from tangent bundle geometry for scalar fields on flat spacetime $(M,\eta)$, where we use the metric signature convention $(-,+,+,+)$. A real scalar field is a function $\phi:M\rightarrow\mathbb{R}$ on the manifold~$M$, and the free field equation in a global Cartesian coordinate system is
\begin{equation}\label{eqn_freeflat}
(\eta^{ab}\partial_a\partial_b -m^2)\phi(x)=0\,.
\end{equation} 

We observe that the field $\phi(x)$ can be regarded as the field seen from the perspective of the observer at spacetime position $x^a=0$. (This becomes clear by thinking of observers in the following general relativistic way: to describe physics, each observer uses an orthonormal frame along his worldline. All tensor fields are then interpreted with respect to this frame. In flat spacetime, at a given worldline position, global coordinates $(x^a)$ centered at this position would be preferred by the observer, so that the frame vectors coincide with the corresponding partial derivatives $\partial_a$.) This point of view leads us to reinterpret $\phi(x)$ as $\Phi(0,x)$ where the coordinates $(x^a)$ now encode the vector pointing to where the field is observed from the origin. For later convenience we use vector coordinates $(u^a)$ from now on. The standard scalar field formulation for $\phi:M\rightarrow\mathbb{R}$ on flat spacetime thus becomes fully equivalent to a formulation in terms of 
$\Phi_x: T_xM\rightarrow\mathbb{R}$ defined by $\Phi_x(u)=\Phi(0,u)=\phi(u)$ as a function on some observer's tangent space $T_xM$ at $x$ with centered coordinates $x^a=0$. The equation of motion now is written in terms of $u$-derivatives $\partial_{\bar a}=\partial/\partial u^a$ as $(\eta^{ab}\partial_{\bar a}\partial_{\bar b} -m^2)\Phi(0,u)=0$. 

Our procedure, at this point, appears to be no more than a naive relabelling of the coordinates $x$ of $M$ into coordinates $u$ of some $T_xM$. But we arrived there by a subtle shift in the interpretation: $\Phi_{x=0}(u)=\Phi(0,u)$ is the field seen at $x$ in the direction provided by coordinates $u$ on $T_xM$. Importantly, this notion can be taken one step further --- by defining $\Phi$ simultaneously for all possible observer positions~$x$ on spacetime, i.e., on the tangent bundle $TM$. Then a scalar field becomes a function $\Phi:TM\rightarrow\mathbb{R}$ satisfying the tangent bundle equations
\begin{eqnarray}
\big(\eta^{ab}\partial_{\bar a}\partial_{\bar b}-m^2\big) \Phi(x,u) &=& 0\,,\label{eqn_flatfield}\\
\big(\partial_a - \partial_{\bar a}\big) \Phi(x,u) &=& 0\,.\label{eqn_flatconsistency}
\end{eqnarray}
The first is the equation of motion, while the second equation can be regarded as a consistency condition, ensuring that the degrees of freedom of the tangent bundle function $\Phi$ are restricted to those of a scalar field function $\phi$ on spacetime.

The new scalar field formulation using equations (\ref{eqn_flatfield}) and (\ref{eqn_flatconsistency}) is still equivalent to the standard one. To see this, note that, in local coordinates, we may introduce new variables $z^{a\pm}=x^a\pm u^a$ so that the spacetime field condition becomes $2\partial/\partial_{z^{a-}}\Phi=0$ and guarantees solutions of the form $\Phi(x,u)=\phi(x+u)$ for some spacetime function $\phi$. Hence the full tangent bundle function $\Phi(x,u)$, when restricted to a single fibre of the tangent bundle at a point with coordinates $(x^a)$, precisely encodes the scalar field as seen by an observer at this point in direction $u$. Observers at other spacetime points can follow the same procedure. Hence our tangent bundle formulation of flat spacetime scalar field theory simultaneously encodes the different perspectives of multiple observers at different spacetime points. We will now make these ideas coordinate-independent and transport them to general curved spacetimes.

\subsection{Generalized field equation}\label{sec_genfieldeq}
In this subsection, we will show that there exists a covariant generalization of equation~(\ref{eqn_flatfield}) to curved spacetimes $(M,g)$, despite the unusual appearance of tangent space derivatives. In the following subsection, we will then discuss the generalization of the spacetime field condition (\ref{eqn_flatconsistency}).

As explained above, the scalar field seen by an observer at point $x$ is the restriction of the tangent bundle field $\Phi$ to the fibre $T_xM$. In induced coordinates, we may write this restriction as  $\Phi_x(u)=\Phi(x,u)$. In close analogy to the standard free scalar field action, such an observer at worldline position $x=x(\tau)$ would describe physics by the action
\begin{eqnarray}\label{eqn_action}
S[\Phi_x] &=& \int_{T_xM}d^4u\sqrt{-g}\frac{1}{2}\Big(g^{ab}\partial_{\bar a}\Phi_x\partial_{\bar b}\Phi_x + m^2 \Phi_x^2\Big) \nonumber \\
&=& \int_{T_xM}d^4u\sqrt{-g}\frac{1}{2} \Big( (g^{-1})^{V\,AB} \partial_A\Phi\partial_B\Phi +m^2\Phi^2\Big).
\end{eqnarray}
This action depends on $x(\tau)$ via the field restriction $\Phi_x$ and the metric $g$. Though the metric integration measure is independent of the tangent space coordinates $(u^a)$, it is included in order to guarantee that $S[\Phi_x]$ is a function on the spacetime manifold $M$. The second line above uses tangent bundle notation, namely, induced coordinates and the vertical lift of the inverse metric, see appendix~\ref{app_basic}.

The variation of $S[\Phi_x]$ with respect to $\Phi_x$ generates the free field equation
\begin{equation}
\big(g^{ab}(x)\partial_{\bar a}\partial_{\bar b}-m^2\big) \Phi_x(u)=0\,,
\end{equation}
which correctly simplifies to~(\ref{eqn_flatfield}) in the flat spacetime case. The equation is also fully covariant on the tangent bundle. Indeed, it can be rewritten in the form
\begin{equation}\label{eqn_field}
\Big( (g^{-1})^{V\,AB}\nabla_A\partial_B - m^2\Big) \Phi(x,u)=0
\end{equation}
provided that the coefficients $\Gamma^{A}{}_{BC}$ of the tangent bundle linear connection $\nabla$ used here satisfy $\Gamma^A{}_{\bar b\bar c}=0$ in induced coordinates. Linear connections of this type do indeed exist, for example the Levi-Civita connections of metrics II or I+II discussed in~\cite{Yanobook}.

A tangent bundle coordinate system that is particularly well-adapted to the perspectives of physical observers is defined in~(\ref{eqn_obscoords}). In these observer coordinates~$(x^\mathcal{A})$, the field equation (\ref{eqn_field}) becomes particularly simple. Making use of the observers' frames, the vertical lift of the inverse metric $(g^{-1})^V$ then is simply determined by the Minkowski metric, see~(\ref{eqn_metriclift}). Moreover, using $\Gamma^A{}_{\bar b\bar c}=0$, the transformation formula for connection coefficients under changes of coordinates implies that also ${\Gamma^\mathcal{A}{}_{\hat\mu\hat\nu}=0}$. Combining these observations, the field equation in observer coordinates hence takes the form
\begin{equation}\label{eqn_fieldobs}
\big(\eta^{\mu\nu}\partial_{\hat\mu}\partial_{\hat v} - m^2 \big) \Phi(x,u) = 0\,.
\end{equation}
As explained in appendix \ref{app_basic}, a congruence of observers moving over a chart $U\subset M$, makes this equation well-defined on $\pi^{-1}(U)\subset TM$, i.e., on all tangent spaces attached to $U$. 

The curved spacetime field equation formulated on the tangent bundle as in (\ref{eqn_fieldobs}) now has the well-known appearance of the free field equation (\ref{eqn_freeflat}) on flat spacetime. Hence, our tangent bundle formalism will enable us  to perform a general Fourier mode expansion for any observer's perceived free field $\Phi_{x(\tau)}$. Below, we will use this result as the starting point for quantization. 

\subsection{Spacetime field condition}\label{sec_sfcondition}
The spacetime field condition (\ref{eqn_flatconsistency}) restricts the tangent bundle field $\Phi$ so that it effectively becomes a field on spacetime. For flat spacetimes, we observe that fibre derivatives $\partial_{\bar a}$ acting on~$\Phi$ are simply exchanged by the corresponding partial derivatives $\partial_a$ tangent to spacetime.  

This point of view may be generalized to curved spacetimes. The tangent space derivatives are the vertical lifts $\partial_{\bar a}=(\partial_a)^V$ and transform nicely as d-vector fields. The notion of a derivative tangent to spacetime has a standard counterpart in the tangent bundle framework, namely, in the form of a horizontal derivative $(\partial_a)^H=\partial_a-u^p\Gamma^q{}_{ap}(x)\partial_{\bar q}$. As discussed in appendix~\ref{app_basic}, the definition of horizontality requires a connection, which we take to be the standard Levi-Civita connection. Also the horizontal derivative transforms as a d-vector field. With these generalizations at hand, we would like to implement, for all $a=1,\dots, 4$, the following condition in order to replace (\ref{eqn_flatconsistency}),
\begin{equation}\label{eqn_tryconsistency}
Z_a \Phi(x,u)=\big((\partial_a)^H - (\partial_a)^V \big) \Phi(x,u) = 0\,.
\end{equation}

However, calculating the commutator of the vector fields $Z_a$ yields the result
\begin{equation}
[Z_c,Z_d]^A = \left[\begin{array}{c}0\\-R^a{}_{pcd}(x)u^p\end{array}\right] = -R^a{}_{pcd}(x)u^p(\partial_a)^{V\,A}\,.
\end{equation}
So the $Z_a$ do not form a subalgebra of the vector fields over the tangent bundle unless the Riemann curvature of spacetime vanishes. The Frobenius theorem for differential equations~\cite{bookdiffeq} then implies that there do not exist local solutions of the condition (\ref{eqn_tryconsistency}) for the field $\Phi$ except on flat spacetime. That the condition works in the flat scenario is what we have already seen in section~\ref{sec_flatmotivation}.
Unless the Riemann tensor has very special properties, thus allowing for a commuting subset of the fields~$Z_a$, the best restriction that still allows local solutions for $\Phi$ on general curved spacetimes is given by the action of a linear combination, 
\begin{equation}\label{eqn_try2}
T^a(x,u)Z_a \Phi(x,u)=0\,.
\end{equation}
Note that a condition in this form does not sufficiently reduce the degrees of freedom of $\Phi$; locally, it would still depend on seven suitable coordinates, not on four, as required for a spacetime field. Moreover, it would not be clear which coefficients $T^a$ for the linear combination should be chosen. 

A solution for both these issues simultaneously is provided by emphasizing once again the special role of observers in our construction. Any observer interprets the tangent bundle field in the form of $\Phi(x(\tau),u)$ along their worldline, only. The observer's four-velocity is a preferred vector field along the worldline, so it is natural to choose $T^a=e_0^a$, and condition (\ref{eqn_try2}) becomes
\begin{equation}\label{eqn_consistency}
e_0^a(\tau)Z_a \Phi(x(\tau),u)=0\,.
\end{equation}
Using a similar counting argument as above, this condition restricts $\Phi$ locally to depend only on four suitable variables. In this sense, each observer may then interpret $\Phi$ as a field on spacetime.

We will now provide another geometric interpretation, and justification, of our observer-dependent generalization~(\ref{eqn_consistency}) of the spacetime field condition (\ref{eqn_flatconsistency}) from flat to curved spacetimes. We first expand the condition in components, then we use a normal coordinate system at position $x^a(\tau)$ on the observer's worldline, where $\Gamma^q{}_{ap}(x(\tau))=0$, and ${e_0^a(\tau)=\dot x^a(\tau)}$ to arrive at
\begin{eqnarray}
0 &=& e_0^a(\tau)\big(\partial_a-\partial_{\bar a} - u^p\Gamma^q{}_{ap}(x(\tau))\partial_{\bar q}\big) \Phi(x(\tau),u) \nonumber\\
&=& \big(\partial_\tau - e_0^a(\tau)\partial_{\bar a}\big)\Phi(x(\tau),u)\,.
\end{eqnarray}
Observe, that this is solved by $\Phi(x(\tau),u)=\phi(x(\tau) + u)$ for some function $\phi$. Since geodesics through $x(\tau)$ are linear functions in their parameter in our normal coordinate system, we find that $x(\tau) + u$ is the position of the point at parameter distance one on the geodesic through~$x(\tau)$ in tangent direction $u$. Hence, using the exponential map, we may write
\begin{equation}\label{eqn_globalcon}
\Phi(x(\tau),u)=\phi(\exp_{x(\tau)}u)
\end{equation}
in a form that holds in all induced coordinate systems on the tangent bundle.

\begin{figure}[ht]
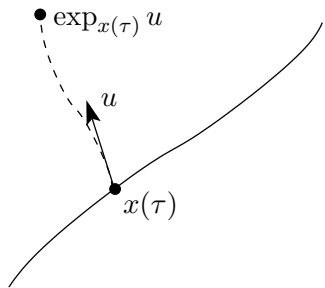
\caption{Illustration of the spacetime field condition: the tangent bundle field $\Phi(x(\tau),u)$ observed in direction $u$ from $x(\tau)$ is determined by a spacetime function $\phi(\exp_{x(\tau)}u)$, evaluated at a point on the geodesic starting at $x(\tau)$ in direction $u$. \label{fig_map}}
\end{figure}

The picture behind this is as follows. A scalar field in the tangent bundle formulation is, from an observer's perspective, determined by a function on spacetime. This results from the spacetime field condition (\ref{eqn_consistency}) which we have seen appropriately to restrict the information content of the tangent bundle field. The spacetime field condition is illustrated in figure~\ref{fig_map}.

\subsection{Standard versus tangent bundle formalism}\label{sec_comparison}
We now come to an important difference between the flat and the curved spacetime case. The classical flat spacetime field theories are fully equivalent in our tangent bundle formulation and in the standard formulation. Indeed, once the spacetime field condition (\ref{eqn_flatconsistency}) is implemented, the equation of motion (\ref{eqn_flatfield}) takes the standard form and leads to the same set of solutions. In the case of general curved spacetimes, we have seen that the direct generalization (\ref{eqn_tryconsistency}) of the spacetime field condition could not be implemented because of the Frobenius theorem. The successful modification~(\ref{eqn_consistency}), however, depends on the perspective of a given observer. Although every observer will see the tangent bundle field $\Phi$ determined by a spacetime field $\phi$ according to (\ref{eqn_globalcon}), this field will generally not agree for different observers, but depend on their worldline.

To conclude, our new tangent bundle formulation of classical field theory on flat spacetime appears to be fully equivalent to the standard formulation. But for general curved spacetimes the tangent bundle construction yields a new, non-equivalent, generalization of flat spacetime field theory. In a sense, our construction provides a highly non-trivial non-minimal coupling procedure of fields to the spacetime metric. We will see below that the quantum theory based on our construction will have different properties than in the standard formulation, and that this will also apply to certain aspects of flat spacetime quantum field theory.

\section{Canonical quantization along observer worldlines}\label{sec_quantization}
We will now describe a two-step procedure to derive the quantum theory corresponding to our tangent bundle formulation of classical field theory. First, in section~\ref{sec_modeexp}, we will perform canonical quantization for the solutions $\Phi_x$ of the field equation (\ref{eqn_field}) at every point $x$ in spacetime. In the following section~\ref{sec_timepart} we will then establish the spacetime field condition (\ref{eqn_consistency}) for the quantum fields $\Phi_{x(\tau)}$ seen by each observer at different points $x(\tau)$ along their worldline in spacetime. For the classical field, we already know that this led to a reduction of the degrees of freedom of the tangent bundle field $\Phi$ so that, effectively, it becomes a spacetime function from an observer's perspective. For the quantum field, we will establish the result that the spacetime field condition guarantees a consistent particle interpretation.

\subsection{Mode expansions and scalar products}\label{sec_modeexp}
The starting point for quantization is the simple form~(\ref{eqn_fieldobs}) that the free scalar field equation~(\ref{eqn_field}) takes in observer coordinates $(x^\mathcal{A})$ on $TM$. Since this has the same form as it would have on flat spacetime in the usual geometric framework, we may proceed by canonical quantization, see e.g. \cite{Kakubook}. It is worth emphasizing, however, that the tangent bundle construction puts us into the position to work out a quantization of the field on a general curved Lorentzian manifold $(M,g)$.

In a given observer coordinate system at $x\in M$, the solution $\Phi_x$ of equation~(\ref{eqn_fieldobs}) has a standard Fourier expansion 
\begin{equation}\label{eqn_expansion}
\Phi(x,u) = \int \frac{d^4p}{(2\pi)^3}\,\delta(\eta^{\mu\nu}p_\mu p_\nu+m^2)\theta(p_{\hat 0}) \Big( A(x,p)e^{-ip_\mu u^\mu}+ A(x,p)^*e^{ip_\mu u^\mu}\Big);
\end{equation}
the dependence on the observer's position enters only into the Fourier coefficients. The step function chooses the future lightcone, and so does not break Lorentz invariance. Integrating out $p_{\hat 0}$, and replacing the Fourier coefficients by creation and annihilation operators yields
\begin{equation}\label{Phimodes}
\Phi(x,u)= \int \frac{d^3p}{(2\pi)^3\,2p_{\hat 0}} \Big(a(x,\bm p) e^{- i p_\mu u^\mu}+a^\dagger(x,\bm p) e^{ip_\mu u^\mu}\Big)
\end{equation}
where $p_{\hat 0} = \sqrt{\bm p^2+m^2}$ abbreviates positive energy. The integration measure and the exponents are locally Lorentz-invariant (though not manifestly so) under coordinate changes between different sets of observer coordinates. More covariance is not required, since the mode decompositions are defined by each observer with respect to their observer coordinates. Note that the momentum dependence of the operators arises from the original Fourier coefficient functions on the cotangent  bundle, as, for instance, in $a(x,\bm p)=A(x,p_{\hat 0}(\bm p),\bm p)$.  

For functions on the tangent bundle fibre $T_xM$, a standard scalar product~\cite{BirrellDavies} can be defined by integration along a hypersurface of constant perceived time $u^{\hat 0}$ in observer coordinates, 
\begin{equation}
\big( \Phi_1(x,u), \Phi_2(x,u) \big) = -i \int d^3u\, \Phi_1^*(x,u)\!\stackrel{\leftrightarrow}{\partial}_{\hat 0} \Phi_2(x,u)\,.
\end{equation}
This definition yields the following orthogonality conditions for the Fourier modes,
\begin{equation}\label{eqn_orthogonality}
\big( e^{\pm ip_\mu u^\mu}, e^{\pm iq_\mu u^\mu}\big) = \pm (2\pi)^3 2p_{\hat 0} \delta^3_{\bm p}(\bm q)\,,\qquad
\big( e^{\mp ip_\mu u^\mu}, e^{\pm iq_\mu u^\mu}\big) = 0\,,
\end{equation} 
which in turn can be used to solve~(\ref{Phimodes}) for the particle creation and annihilation operators:
\begin{equation}\label{eqn_operators}
a(x,\bm p)= -\big( e^{- i p_\mu u^\mu}, \Phi(x,u) \big)\,, \qquad
a^\dagger(x,\bm p)= \big( e^{i p_\mu u^\mu}, \Phi(x,u) \big)\,.
\end{equation}

Note that each of the Lorentz-symmetry related observers at position $x$ in spacetime finds the same set of particle operators $a(x,\bm p)$ and $a^\dagger(x,\bm p)$, as it is the case for standard flat spacetime quantum field theory. Hence all observers at $x$ can agree on a common definition of the vacuum state by
\begin{equation}
a(x,\bm p) \left| 0 \right> =0 \quad \forall \bm p\,.
\end{equation}
Without further arguments, this vacuum state could still depend on the observer position. However, the particle observation theorem of the following section will show that the vacuum state is transported consistently along the observers' worldlines; consequently, $\left| 0 \right>$ cannot have a spacetime-dependence. Based on the vacuum definition a Fock space can be constructed. Despite the fact that this Fock space is the same for each observer, our discussion in section~\ref{sec_infoenergy} will show that each observer must store his information in their own state. In this way our construction enables us to model different sets of information, perhaps resulting from different measurements.

\subsection{Consistency of particle measurements}\label{sec_timepart}
We will now implement the spacetime field condition (\ref{eqn_consistency}) acting on the quantum field. We will see that this condition allows us to identify the particle creation and annihilation operators under time-evolution along an observer's worldline. In consequence a particle picture will emerge, where quantum particle momentum states are observed in a way that is consistent with the general relativistic arguments of parallel transport.

Since the quantum field in our construction has the mode expansion (\ref{Phimodes}) in observer coordinates, we need to transform the spacetime field condition into the same coordinate system. First, consider the vector fields
\begin{equation}
Z_a =(\partial_a)^H - (\partial_a)^V = \partial_a - u^p\Gamma^q{}_{ap}(x)\partial_{\bar q} - \partial_{\bar a}
\end{equation}
in induced coordinates of $TM$. The transformation of the partial derivatives to observer coordinates is displayed in (\ref{eqn_indtoobs}) and leads to 
\begin{equation}
Z_a = \partial_a + u^p \partial_a e^\mu_p \partial_{\hat \mu} - u^p\Gamma^q{}_{ap}(x) e^\mu_q \partial_{\hat \mu} - e^\mu_a\partial_{\hat \mu}
= \partial_a + u^\rho e^p_\rho \nabla_a e^\mu_p \partial_{\hat \mu} - e^\mu_a\partial_{\hat \mu}\,,
\end{equation}
where the standard Levi-Civita connection appears. Second, we linearly combine the fields $Z_a$ with the observer's four-velocity along their worldline,
\begin{eqnarray}
e_0^a(\tau) Z_a &=& \dot x^a\partial_a + u^\rho e^p_\rho \nabla_{e_0} e^\mu_p \partial_{\hat \mu} - \partial_{\hat 0}\nonumber\\
&=& \dot x^a(\tau)\partial_a - \Omega^\mu{}_\rho(\tau) u^\rho \partial_{\hat \mu} - \partial_{\hat 0}\,,
\end{eqnarray}
where we employed the duality relation $e^p_\rho e^\mu_p=\delta^\mu_\rho$, and inserted the matrix $\Omega^\mu{}_\rho$ of covariant acceleration and rotation for an arbitrary observer, making use of its definition $\nabla_{e_0}e_\rho=\Omega^\nu{}_\rho e_\nu$. The dependence on the worldline parameter is explicitly shown in the final expression; note that~$u^\rho$ are the observer coordinate components of a general tangent space vector at $x(\tau)$.

Finally, we may act on the quantum field along the observer's worldline, see (\ref{eqn_consistency}), to find 
\begin{eqnarray}\label{eqn_qucon}
0 &=& e_0^a(\tau) Z_a\Phi(x(\tau),u) \nonumber\\
&=& \int \frac{d^3p}{(2\pi)^3\,2p_{\hat 0}} \Big(\big(\dot x^a\partial_a  +i \Omega^\mu{}_\rho(\tau) u^\rho p_\mu +i p_{\hat 0}\big) a(x,\bm p)  e^{- i p_\mu u^\mu} + \mathrm{H.c.} \Big)
\end{eqnarray}
where the Hermitian conjugate part, as well as the worldline parameter dependence for $x(\tau)$,~$\dot x(\tau)$, have been suppressed. 

We will first solve this condition for inertial observers for which the covariant acceleration/ rotation matrix~$\Omega$ vanishes. In this case, taking the scalar product with the modes~$e^{\pm i p_\mu u^\mu}$, and using the orthogonality relations (\ref{eqn_orthogonality}), isolates the brackets acting on the particle operators, so that, e.g.,
\begin{equation}
\big(\dot x^a\partial_a +i p_{\hat 0}\big) a(x,\bm p) = 0\,,
\end{equation}
which condition is solved by 
\begin{equation}\label{eqn_opsin}
a(x(\tau),\bm p) = e^{-ip_{\hat 0}(\tau-\tau_0)}a(x(\tau_0),\bm p)\,.
\end{equation}
For $a^\dagger(x(\tau),\bm p)$ one obtains the Hermitian conjugate. These results reflect how the particle operators of inertial observers evolve in proper-time, or how they are identified along the worldline: In observer coordinates, they obey the usual time-evolution generated by the energy $p_{\hat 0}$. 

Remarkably, this result holds for general curved spacetimes. To interpret it further, we need to consider how momenta change under parallel transport in observer coordinates. We may write
\begin{equation}
0 = \nabla_{e_0} p = \nabla_{e_0} (p_\mu e^\mu) = \big( \partial_\tau p_\mu  -  \Omega^\rho{}_\mu p_\rho\big) e^\mu
\end{equation}
In particular, the case $\Omega=0$ shows that the momentum components in inertial observer coordinates stay constant in proper time.
Hence the $a(x(\tau),\bm p)$ in equation (\ref{eqn_opsin}) are operators for parallely transported momenta at worldline position $x(\tau)$. Since parallel transport identifies the same momenta at different spacetime positions, this means that an inertial observer consistently sees the same vacuum state along his worldline; also the quantum particle momenta measured at some initial position $x(\tau_0)$ will agree with those measured at a later position $x(\tau)$. 

This discussion can be generalized for observers with arbitrary covariant rotation and acceleration; in appendix \ref{app_proof}, we will prove the following much stronger result.

\vspace*{3pt}\noindent 
\textit{\textbf{Particle observation theorem.} The particle operators $a(x(\tau),{\bm p}(\tau))$ at worldline position $x(\tau)$ for parallely transported momentum ${\bm p}(\tau)$ are obtained from those at earlier proper time $\tau_0$ by the time-evolution formula
\begin{equation}\label{eqn_timeevo}
a(x(\tau),{\bm p}(\tau)) = T\exp \Big(- i \int_{\tau_0}^\tau d\tau'\, p_{\hat 0}(\tau')\Big) a(x(\tau_0),{\bm p}(\tau_0))
\end{equation}
where $T$ denotes the time-ordering of the integration variables in the exponential power series. The creation operators obey the Hermitian conjugate equation.}
\vspace*{3pt}

The formula above identifies the particle operators at different worldline positions. It shows a structural agreement with the Heisenberg time-evolution where the relevant Hamiltonian is simply given by the particle energy in the observer's tangent bundle coordinate system. 
Since parallel transport in general relativity physically identifies the same momenta at different spacetime points (thinking of gyroscopic measuring), the particle observation theorem implies that each observer of a non-interating quantum field measures an initial vacuum state or momentum state completely unchanged along their worldline at later times.

To conclude, the spacetime field condition acting on the quantum field guarantees a consistent time-evolution of the vacuum and of particle momentum states, and so provides a consistent particle interpretation for any observer on a general curved spacetime.

\section{Quantum information, particles and energy}\label{sec_infoenergy}
We will now take a look at some consequences resulting from our new quantum field theory construction. We will find that the present theory enforces a strong link between the concepts of information, particle observation and energy-momentum.

\begin{figure}[ht]
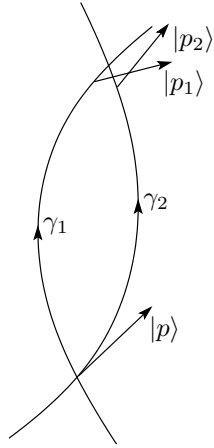
\caption{Consequence of the particle observation theorem: if two observers on worldlines $\gamma_1$ and $\gamma_2$ initially agree on the momentum of a measured particle with respect to their local Lorentz frame, then the path-dependence of the ``parallel transport of states'' will lead to their disagreement at a later meeting point due to spacetime curvature.\label{fig_statetransport}}
\end{figure}

Consider the following idealized situation of two communicating observers. They meet at the same spacetime point and  agree there on a measured particle state, say a single-particle state of a certain momentum $\left| p \right>$, appropriately Lorentz transformed from one observer's frame to the other's; the latter will always be assumed in the remainder of this section. Now both observers move on in time along their worldlines and meet at a later spacetime point. Since the momentum the observers measure is parallely transported along their worldlines, as we have seen in section~\ref{sec_timepart}, both would say that their particle momentum stays unchanged.  But there will be a mismatch when the observers meet again, due to the path-dependence of parallel transport, which is measured by the curvature of spacetime. For non-vanishing curvature, the observers will no longer agree on their respective particle momenta $\left| p_1 \right>$ and $\left| p_2 \right>$, only on having a single-particle state, see figure~\ref{fig_statetransport}.

This apparent contradiction in the theory is based on the standard assumption that there is a single state of the quantum field measured by both observers. It is easily resolved by attaching a state to each observer, thereby facilitating their own description of the physical world. This point of view is also advantageous since it easily allows the description of situations where observers have different information, for instance, because of different measurement methods. 

Moreover, there is another, very interesting consequence. For the situation described above, and in figure~\ref{fig_statetransport}, the initial measurement where the observers agree means they use their own states $\left| p \right>_1$ and $\left| p \right>_2$ of the same momentum in the respective frames; we interpret this by saying both are convinced to observe a particle of momentum $p$. When they meet again, several outcomes are possible:
\begin{enumerate}[$\;\bullet\;$]\itemsep-3pt
\item If the observers do not exchange information, they will move on along their worldlines, still being convinced of their respective single particle states  $\left| p_1 \right>_1$ and $\left| p_2 \right>_2$ of parallely transported momenta $p_1$ and $p_2$, totally unchanged, if compared to their gyroscopic measuring devices. 
\item One of the observers, say observer 1, could communicate his momentum measurement $\left| p_1 \right>_1$ without receiving any information in return. Then 1 would move on convinced of his current unchanged particle state $\left| p_1 \right>_1$. However, observer 2 would have to take into account the information obtained (assuming no reason to distrust 1) and would move on, now being convinced of a two-particle state $\left| p_1 p_2 \right>_2$. This state arises from the unchanged momentum observation of observer~2 and from the communicated momentum $p_1$ that differs by a curvature contribution.
\item If the observers both communicate their measurements, they will move along now both being convinced of a two-particle state, i.e., $\left| p_1 p_2 \right>_1$ or $\left| p_1 p_2 \right>_2$.
\end{enumerate}
We remark that the latter two scenarios are not time-reversible; here \textit{communication is responsible for the arrow of time}.
As a second remark, observe that already in standard quantum field theory, measuring particles or receiving a communication about a measurement, and thus gathering information, results in the description of the quantum world by means of certain particle states, so one could say that information creates particles. Our analysis above, in our new quantum field theory formalism may then be summarized by the  following statement: \textit{Information creates additional particles on curved spacetimes}.

This idea can be reinforced by speculating on the gravitational coupling of our theory. Supposing that the Einstein equations should be a good limiting case of our theory, the gravitational equations of motion should take the form
\begin{equation}
R^{ab}(x)-\frac{1}{2}R(x)g^{ab} = \sum_{\mathrm{observers}\;i} \int_{\gamma_i} d\tau \frac{\delta^4_{x}(x(\tau))}{\sqrt{-g(x(\tau))}}T^{ab}(x(\tau),0)\,.
\end{equation}
The total energy-momentum should be a sum over all observers with worldlines $\gamma_i$ contributing equally, because each observer uses the action~(\ref{eqn_action}) to describe the field. In case the observers move along a congruence of worldlines in some neighbourhood, the complete matter side of the equation would reduce to a single energy-momentum tensor $T^{ab}(x)$, so providing the required Einstein limit. The separate energy-momentum contributions in the equation could arise from~(\ref{eqn_action}) by variation with respect to the spacetime metric, and by applying the spacetime-field condition. We expect that this leads to the standard form of the energy-momentum tensor. The fact that the evaluation takes place at $u=0$ can be interpreted by saying that each observer will only contribute energy-momentum localized on his worldline; indeed this is required to prevent an infinite overcounting of energy-momentum contributions in the Einstein limit.

Of course the proposed gravitational coupling will have to be analysed more carefully. But let us take it serious in the context of our information arguments, and in a semiclassical setting, where the spacetime metric is classical, but coupled to quantum field theory in our formalism. Since each observer has access to physical information stored in his quantum state, and will use it to contribute an expectation value of his energy-momentum contribution, we could conclude this section by claiming: \textit{Information generates energy momentum and thereby changes spacetime}.

\section{Conclusion and outlook}\label{sec_discuss}
In quantum field theory a particle state of definite momentum is a non-local concept. On this basis we have argued in general relativistic fashion that an observer on spacetime who measures such a particle initially would expect to measure the same momentum later, where initial and final momentum measurement are identified by parallel transport along the observer's worldline. However, this line of thought is not realized by standard quantum field theory on curved spacetime, where the particle concept suffers an almost complete breakdown because it appears to be tied strongly to the flatness of spacetime and unbroken Poincare symmetry.

In this paper we have used tangent bundle geometry to generalize classical field theory on flat spacetime, finally arriving at a new construction of quantum field theory on general curved spacetimes. In a sense, our construction realizes a non-trivial non-minimal coupling scheme of fields to the spacetime metric. 

One of the great benefits of our framework is that the field equations are now interpreted to live on the flat tangent spaces where, as is well-known, the Lorentz group acts on observers' frames. Canonical quantization hence becomes possible in suitable observer-adapted coordinates. The eight-dimensionality of the tangent bundle field could be reduced by establishing what we called the spacetime field condition, and led us to an interpretation of the field theory in terms of multiple observers' perspectives; a tangent bundle field $\Phi(x,u)$ corresponds to a spacetime field seen by an observer at spacetime position $x$ when he looks into the tangent direction $u$. In this way, our construction provides a geometric basis for the interpretation of the Poincare group action on the local tangent spaces. 

For the quantum theory, we could prove the particle observation theorem around equation~(\ref{eqn_timeevo}). It states that the creation and annihilation operators of particle momentum states along an observer's worldline can be calculated by a  Heisenberg-type time-evolution from the respective operators at earlier proper times, without their mixing. Placatively, the particle momentum states are parallely propagated along each observer's worldline. Thus, our construction realizes a particle picture in quantum field theory on curved spacetimes which is nicely consistent with general relativistic reasoning. It follows in particular that a preferred vacuum state can be consistently defined by all observers.

We have presented arguments for the need to attach a state to each observer in the theory. In consequence we have seen that information and energy in our formalism become closely linked:

On the one hand, information can result in the creation of additional particles on curved spacetimes, while this effect does not arise on flat spacetimes. Combined with the implication of our particle observation theorem, i.e., that creation and annihilation operators do not mix from any observer's perspective, this seems to leave little room for the Davies-Unruh effect within our theoretical framework. But this certainly has to be inspected more carefully in a future publication, as well as Hawking's result of particle radiation by black holes. In order to do this, the interpretation of observers and their role in our theoretical framework should be further clarified, since this has a strong influence on applications.

On the other hand, our proposal of the form of a semi-classical gravitational coupling of the quantum fields to the spacetime metric, suggests that additional observers increase the amount of energy-momentum that creates spacetime. 
Several possible consequences of this idea present themselves: In the double-slit experiment the observation of path-information, i.e., the presence of an additional observer, leads to decoherence, and destroys the interference pattern\cite{Herzog:1995zz}; this effect might now be explained as being due to the extra curvature, and hence the enhanced path-dependence discrepancy between the different particle trajectories. Such a scenario could be of additional interest, also in the discussion of interaction-free measurements, as in the prototype of the bomb-testing problem~\cite{Elitzur:1993xh}: The described process of decoherence due to an additional observer does not require the observer to communicate, it simply requires their ability to communicate. A quantum eraser~\cite{Walborn:2002zza} might then be interpreted as removing such an able observer from the theory. Whether the magnitude of curvature contributions would be sufficiently large to cause decoherence in the experiments mentioned, remains to be analysed. What we have suggested here is a new form of decoherence; it is not strictly gravitational as suggested by various other approaches, see the review~\cite{Bassi:2017szd}; rather, it is based on the presence and communiction of information and merely mediated by gravity.

In our framework the quantum regime seems to be realized by experimental situations with few observers sparsely distributed in spacetime. The presence of more and more observers, for instance in the form of objects that scatter photons and thereby transmit information, leads to decoherence by changing the curvature of spacetime and thus increasing the path-dependence of quantum state transport. The special limiting case of infinitely many observers along a congruence of curves would provide the classical Einstein coupling of matter to spacetime. This conclusion becomes possible without access to a full quantum gravity theory; it is based solely on our new generalization of classical flat spacetime field theory to quantum field theory on curved spacetimes.

\acknowledgements
MNRW is very grateful to Felix Tennie for fruitful discussions and to Georg von Hippel for useful comments.

\appendix
\section{Basic tangent bundle geometry}\label{app_basic}
In this appendix, we review some of the geometric notions required in this paper. In particular, we shall display some very useful coordinate systems for the tangent bundle $TM$ to spacetime and make a few remarks about its differential geometry. 

The basic spacetime geometry plays on the stage of a Lorentzian manifold $(M,g)$ with a metric of signature $(-,+,+,+)$. When we speak about an observer, they are characterized by their worldline, i.e., by some timelike curve $\gamma$ on $M$ with worldline parameter $\tau$. Moreover each observer carries a frame of reference $\{e_0(\tau)=\dot\gamma(\tau),e_\alpha(\tau)\}$ defined along $\gamma$ which satisfies the orthonormality condition $g(e_\mu,e_\nu)=\eta_{\mu\nu}$. The frame provides a vector basis of the worldline's tangent spaces $T_{\gamma(\tau)}M$. The associated dual bases of $T^*_{\gamma(\tau)}M$ are provided by the orthonormal coframes $e^\mu$ that are defined by the duality condition $e^\mu(e_\rho)=\delta^\mu_\rho$.

\subsection{Coordinate systems on TM}
Each point of the tangent bundle $TM$ of the spacetime manifold $M$ is an element of the fibre $T_xM$ over some point $x\in M$. Using spacetime coordinates $(x^a)$ so provides induced local coordinates for points $u$ of $TM$, as
\begin{equation}
(x^A)=(x^a,u^{\bar a})
\end{equation}
where $u=u^a\partial_a \in T_xM$. Note that $u^{\bar a}=u^a$ carries the usual coordinate index, but the barred index notation is convenient sometimes, e.g., for partial derivatives $\partial_{\bar a}=\frac{\partial}{\partial u^a}$ along the tangent bundle fibres. 

Changing coordinates on $M$ from $x^a$ to $\tilde x(x)^a$ changes the induced coordinates on $TM$,
\begin{equation}
(x^A) \quad \rightarrow \quad (\tilde x^A)  = (\tilde x^a, \tilde u^{\bar a} ) = (\tilde x^a, \frac{\partial\tilde x^a}{\partial x^b} u^b) \,.
\end{equation}
For later convenience, we display the associated transformation of TM-vector components ${\tilde Z^A = C^A{}_B Z^B}$. This is obtained from the relation $d\tilde x^A=C^A{}_B dx^B$ as 
\begin{equation}\label{tmvtrans}
C^A{}_B = \left[\begin{array}{cc}C^a{}_b & C^a{}_{\bar b}\\C^{\bar a}{}_b&C^{\bar a}{}_{\bar b}\end{array}\right] = 
\left[\begin{array}{cc} \frac{\partial\tilde x^a}{\partial x^b} & 0\\  \frac{\partial^2\tilde x^a}{\partial x^b\partial x^c} u^c & \frac{\partial\tilde x^a}{\partial x^b}  \end{array}\right].
\end{equation}

Instead of induced tangent bundle coordinates for $u\in TM$, we may employ some observer's frame to decompose $u=u^\mu e_\mu$ into frame components. This leads us to the definition of an observer-adapted local coordinate system on $TM$, which we shall call \textsl{observer coordinates},
\begin{equation}\label{eqn_obscoords}
(x^\mathcal{A})=(x^a,u^{\hat \mu})\,.
\end{equation}
Similarly as above, $u^{\hat \mu}=u^\mu$ carries the usual frame index, but the hatted indices can be used to conveniently abbreviate the partial derivatives $\partial_{\hat \mu}=\frac{\partial}{\partial u^\mu}$. Moreover, the different index notations allow us to distinguish induced and observer coordinates, e.g., in $u^{\bar 0}$ and $u^{\hat 0}$.

Observer coordinates are initially defined along the worldline $\gamma$ of a single observer (open neighbourhood $\gamma\,\cap U$ where $\gamma$ lies in the chart $U$ of $M$). This means that observer coordinates coordinatize $\pi^{-1}(\gamma)$, i.e., a piece of the worldline with the tangent space fibres of the bundle attached. If we consider a congruence of curves on $U\subset M$, along which certain observers move, then observer coordinates provide more general charts for $TM$, based on $\pi^{-1}(U)$. 

Changing coordinates on $M$ from $x^a$ to $\tilde x(x)^a$ changes the observer coordinates on $TM$ as
\begin{equation}
(x^\mathcal{A}) \quad \rightarrow \quad (\tilde x^\mathcal{A})  = (\tilde x^a, \tilde u^{\hat \mu} ) = (\tilde x^a,u^{\hat \mu}) \,.
\end{equation}
Note that the fibre coordinate is completely unchanged; the reason for this is the coordinate-independent definition of the frames and coframes, and hence of the frame components.

The relation between the two different choices of fibre coordinates, i.e., $u^a$ and $u^{\hat\mu}$, is given by the frames and coframes of the observer. Writing $e_\mu=e_\mu^a\partial_a$ and $\partial_a=e^\mu_a e_\mu$, one finds that $u^\mu=e^\mu_a u^a$ and $u^a=e^a_\mu u^\mu$. It is useful to display the coordinate transformations of the tangent bundle from local induced coordinates to observer coordinates. They are defined by relating the different bases as
\begin{equation}\label{eqn_indtoobs}
dx^\mathcal{A} = M^\mathcal{A}{}_B dx^B\,,\quad \partial_\mathcal{A} = M^{-1\,B}{}_\mathcal{A} \partial_B\,.
\end{equation}
Hence one finds
\begin{equation}
M^\mathcal{A}{}_B = \left[\begin{array}{cc}M^a{}_b & M^a{}_{\bar b}\\M^\mu{}_b&M^\mu{}_{\bar b}\end{array}\right] = \left[\begin{array}{cc}\delta^a_b & 0\\u^p\partial_b e^\mu_p & e^\mu_b \end{array}\right],\quad
M^{-1\,B}{}_\mathcal{A} = \left[\begin{array}{cc}M^{-1\,b}{}_a & M^{-1\,b}{}_\mu\\M^{-1\,\bar b}{}_a & M^{-1\,\bar b}{}_\mu \end{array}\right]
= \left[\begin{array}{cc}\delta^b_a & 0 \\ u^\rho \partial_a e^b_\rho & e^b_\mu\end{array}\right].
\end{equation}
Using the relation $e^\mu_a \partial_c e^a_\rho = - e^a_\rho \partial_c e^\mu_a$, which follows from $e^\mu_a e^a_\rho= \delta^\mu_\rho$ by acting with $\partial_c$, it is easy to check from these expressions that these matrices are inverses of one another.

\subsection{Geometric constructions on $TM$}
Since the equations in our field theory formulation are written on the eight-dimensional manifold~$TM$, we need to consider vectorial derivatives in all eight directions. The vector fields on $TM$ are sections of the bundle $TTM$ that has a decomposition into a horizontal and a vertical part, ${TTM = HTM \oplus VTM}$; compare e.g.~\cite{BucataruMiron} for more background. Here the vertical part is canonically defined, and is spanned by the derivatives along the fibres of $TM$, as 
\begin{equation}
VTM = \langle (\partial_a)^V \rangle =\langle \partial_{\bar a} \rangle\,.
\end{equation}
The horizontal part can be interpreted as being tangent to the spacetime manifold. However, its definition is a matter of some choice, since it requires a connection (that can be non-linear, in general). We simply choose the linear Levi-Civita connection of the spacetime metric to define 
\begin{equation}
HTM=\langle (\partial_a)^H\rangle= \langle \partial_a-u^p\Gamma^q{}_{ap}(x)\partial_{\bar q} \rangle  \,.
\end{equation}
The dual spaces of the horizontal/ vertical decomposition are spanned as follows, 
\begin{equation}
H^*TM= \langle dx^a \rangle\,, \qquad V^*TM=\langle  du^a + u^p\Gamma^a{}_{qp}(x) dx^q \rangle \,.
\end{equation}
The basis elements here can also be written as horizontal and vertical lifts as $(dx^a)^V=dx^a$ and $(dx^a)^H=du^a + u^p\Gamma^a{}_{qp}(x) dx^q$.

Note that the basis vectors and covectors appearing above act locally at points $u$ of the tangent bundle $TM$. Nevertheless, under coordinate changes induced from changes on the spacetime manifold, they transform as standard spacetime vectors and covectors. This fact allows a simple lifting procedure of spacetime tensor fields
\begin{equation}
T=T^{a\dots}{}_{c\dots}\partial_a\otimes\dots\otimes dx^c\otimes\dots
\end{equation}
to tensor fields on the tangent bundle. The resulting dinstinguished tensor fields, or d-tensor fields, on $TM$ also have the property to transform under manifold induced coordinate changes as if they were tensor fields on $M$. Their general form is 
\begin{equation}
T=T^{a\dots b\dots}{}_{c\dots d\dots}(\partial_a)^H\otimes\dots\otimes \partial_{\bar b}\otimes\dots\otimes dx^c\otimes \dots\otimes   (du^d + u^p\Gamma^d{}_{qp}(x) dx^q)  \otimes \dots \,.
\end{equation}
One example of a d-tensor field needed in this paper is the vertical lift of the inverse spacetime metric, i.e.,
\begin{equation}\label{eqn_metriclift}
(g^{-1})^V = g^{ab} \partial_{\bar a}\otimes \partial_{\bar b}  = g^{ab}(e^\mu_a\partial_{\hat\mu})\otimes(e^\nu_b\partial_{\hat\nu}) = \eta^{\mu\nu} \partial_{\hat\mu}\otimes \partial_{\hat\nu}
\end{equation}
in local induced coordinates and observer coordinates on TM, respectively.

\section{Proof of the particle observation theorem}\label{app_proof}
This appendix contains a proof of the particle observation theorem around equation (\ref{eqn_timeevo}) in section \ref{sec_timepart}, which provides a consistent vacuum state and particle interpretation for all observers.

Starting point for the proof is equation (\ref{eqn_qucon}) where the spacetime field condition for a general observer acts on the mode expansion of the free quantum field. For the general observer there appears a term including the worldline dependent covariant acceleration/ rotation matrix $\Omega^\mu{}_\rho(\tau)$ that now cannot be neglected as for the inertial observer. In order to solve condition (\ref{eqn_qucon}) we replace the tangent space coordinates $u^\alpha$ for $\alpha=1,2,3$ by means of 
\begin{equation}
i\frac{\partial}{\partial p_\alpha} e^{-ip_\mu u^\mu} = \Big(u^\alpha + u^{\hat 0}\frac{p^\alpha}{p_{\hat 0}} \Big) e^{-ip_\mu u^\mu}
\end{equation}
where the second term contains $p^\alpha= \delta^{\alpha\beta}p_\beta$, and arises from differentiating $p_{\hat 0}(\bm p)$. This replacement leads to the following contribution to the integrand, which simplifies using the antisymmetry of $\Omega^{\mu\nu}=\Omega^\mu{}_\rho\eta^{\rho\nu}$,
\begin{eqnarray}
&& a(x(\tau),\bm p)\Big(i \Omega^\mu{}_{\hat 0}(\tau)p_\mu u^{\hat 0} - i \Omega^\mu{}_{\alpha}(\tau)p_\mu \frac{p^\alpha}{p_{\hat 0}}u^{\hat 0}  -  \Omega^\mu{}_{\alpha}(\tau)p_\mu \frac{\partial}{\partial p_\alpha}  \Big) e^{-ip_\mu u^\mu} \nonumber \\
&=& {}-  a(x(\tau),\bm p)\Omega^\mu{}_{\alpha}(\tau)p_\mu \frac{\partial}{\partial p_\alpha} e^{-ip_\mu u^\mu}\,. 
\end{eqnarray}
Via integration by parts, the momentum derivatives are then shifted to act on the particle annihilation operator, which gives
\begin{eqnarray}
&& \int \frac{d^3p}{(2\pi)^32p_{\hat 0}}\Big[\Big(- \Omega^\mu{}_{\alpha}(\tau)p_\mu \frac{p^\alpha}{p_{\hat 0}^2} +  \Omega^{\hat 0}{}_{\alpha}(\tau)\frac{p^\alpha}{p_{\hat 0}} +\Omega^\mu{}_{\alpha}(\tau)p_\mu \frac{\partial}{\partial p_\alpha} \Big) a(x(\tau),\bm p)\Big]e^{-ip_\mu u^\mu} \nonumber \\
&& {} - \int \frac{d^3p}{(2\pi)^3} \frac{\partial}{\partial p_\alpha} \Big[\frac{1}{2p_{\hat 0}}\Big(a(x(\tau),\bm p)\Omega^\mu{}_{\alpha}(\tau)p_\mu  e^{-ip_\mu u^\mu}\Big)\Big] \nonumber \\
&=& \int \frac{d^3p}{(2\pi)^32p_{\hat 0}} \Omega^\mu{}_{\alpha}(\tau)p_\mu \frac{\partial}{\partial p_\alpha}a(x(\tau),\bm p)e^{-ip_\mu u^\mu}\,.
\end{eqnarray}
The boundary term at infinite three-momentum can be neglected. (In the algebraic picture the fields are operator-valued distributions tested on functions of compact support.)
The above result for the contributions for non-inertial observers transforms condition (\ref{eqn_qucon}) into
\begin{equation}
0 = \int \frac{d^3p}{(2\pi)^3\,2p_{\hat 0}} \Big[\Big(\dot x^a(\tau)\partial_a + i p_{\hat 0} + \Omega^\mu{}_{\alpha}(\tau)p_\mu \frac{\partial}{\partial p_\alpha}\Big) a(x(\tau),\bm p)  e^{- i p_\mu u^\mu} + \mathrm{H.c.} \Big].
\end{equation}

Via the orthogonality relations (\ref{eqn_orthogonality}) we can extract the condition in round brackets on the particle operators; it is useful to perform the scalar product with modes $e^{\pm i p_\mu(\tau)u^\mu}$, defined in terms of momenta $p_\mu(\tau)$ that are parallely transported along the observers worldline. Then
\begin{eqnarray}
0 &=& \Big(\dot x^a(\tau)\partial_a + i p_{\hat 0}(\tau) + \Omega^\mu{}_{\alpha}(\tau)p_\mu(\tau) \frac{\partial}{\partial p_\alpha}\Big) a(x(\tau),\bm p(\tau)) \nonumber \\
&=& \Big(\frac{d}{d\tau}+ i p_{\hat 0}(\tau)\Big) a(x(\tau),\bm p(\tau)) \,,
\end{eqnarray} 
and the Hermitian conjugate holds for $a^\dagger(x(\tau),\bm p(\tau))$. This equation is solved by the series 
\begin{eqnarray}
a(x(\tau),\bm p(\tau)) &=& \Big(1 +(- i) \int_{\tau_0}^\tau d\tau_1\, p_{\hat 0}(\tau_1) +(- i)^2 \int_{\tau_0}^\tau d\tau_1 \int _{\tau_0}^{\tau_1}d\tau_2 \, p_{\hat 0}(\tau_1)p_{\hat 0}(\tau_2) \\
&& \quad +(- i)^3 \int_{\tau_0}^\tau d\tau_1 \int _{\tau_0}^{\tau_1}d\tau_2 \int_{\tau_0}^{\tau_2} d\tau_3\, p_{\hat 0}(\tau_1)p_{\hat 0}(\tau_2)p_{\hat 0}(\tau_3) + \dots \Big)  a(x(\tau_0),\bm p(\tau_0)) \nonumber
\end{eqnarray}
in which each term's derivative gives the previous term. The whole expression is conveniently abbreviated by the time-ordered exponential as
\begin{equation}
a(x(\tau),{\bm p}(\tau)) = T\exp \Big(- i \int_{\tau_0}^\tau d\tau'\, p_{\hat 0}(\tau')\Big) a(x(\tau_0),{\bm p}(\tau_0))\,.
\end{equation}
Along with the Hermitian conjugate 
\begin{equation}
a^\dagger(x(\tau),{\bm p}(\tau)) = T\exp \Big(+i \int_{\tau_0}^\tau d\tau'\, p_{\hat 0}(\tau')\Big) a^\dagger(x(\tau_0),{\bm p}(\tau_0))\,.
\end{equation}
this is the result we set out to prove in this appendix. The meaning of this result for particle observations by an arbitrarily moving observer is discussed in the main text.


\end{document}

%% file: expmap2.pdf_t
\begin{picture}(0,0)%
\includegraphics{expmap2.pdf}%
\end{picture}%
\setlength{\unitlength}{2171sp}%
\begingroup\makeatletter\ifx\SetFigFont\undefined%
\gdef\SetFigFont#1#2#3#4#5{%
  \reset@font\fontsize{#1}{#2pt}%
  \fontfamily{#3}\fontseries{#4}\fontshape{#5}%
  \selectfont}%
\fi\endgroup%
\begin{picture}(3622,3297)(2367,-5329)
\put(3436,-3241){\makebox(0,0)[lb]{\smash{{\SetFigFont{11}{13.2}{\sfdefault}{\mddefault}{\updefault}{\color[rgb]{0,0,0}$u$}%
}}}}
\put(2896,-2311){\makebox(0,0)[lb]{\smash{{\SetFigFont{11}{13.2}{\sfdefault}{\mddefault}{\updefault}{\color[rgb]{0,0,0}$\exp_{x(\tau)}u$}%
}}}}
\put(3691,-4456){\makebox(0,0)[lb]{\smash{{\SetFigFont{11}{13.2}{\sfdefault}{\mddefault}{\updefault}{\color[rgb]{0,0,0}$x(\tau)$}%
}}}}
\end{picture}%

%% file: statetransport.pdf_t
\begin{picture}(0,0)%
\includegraphics{statetransport.pdf}%
\end{picture}%
\setlength{\unitlength}{1973sp}%
\begingroup\makeatletter\ifx\SetFigFont\undefined%
\gdef\SetFigFont#1#2#3#4#5{%
  \reset@font\fontsize{#1}{#2pt}%
  \fontfamily{#3}\fontseries{#4}\fontshape{#5}%
  \selectfont}%
\fi\endgroup%
\begin{picture}(2098,5594)(3579,-6758)
\put(5653,-1731){\makebox(0,0)[lb]{\smash{{\SetFigFont{10}{12.0}{\familydefault}{\mddefault}{\updefault}{\color[rgb]{0,0,0}$\left| p_2\right>$}%
}}}}
\put(5363,-5381){\makebox(0,0)[lb]{\smash{{\SetFigFont{10}{12.0}{\familydefault}{\mddefault}{\updefault}{\color[rgb]{0,0,0}$\left| p\right>$}%
}}}}
\put(4033,-4031){\makebox(0,0)[lb]{\smash{{\SetFigFont{10}{12.0}{\familydefault}{\mddefault}{\updefault}{\color[rgb]{0,0,0}$\gamma_1$}%
}}}}
\put(5293,-3731){\makebox(0,0)[lb]{\smash{{\SetFigFont{10}{12.0}{\familydefault}{\mddefault}{\updefault}{\color[rgb]{0,0,0}$\gamma_2$}%
}}}}
\put(5533,-2281){\makebox(0,0)[lb]{\smash{{\SetFigFont{10}{12.0}{\familydefault}{\mddefault}{\updefault}{\color[rgb]{0,0,0}$\left| p_1\right>$}%
}}}}
\end{picture}%

%% file: 00_archive_QFT_tangentbundle.bbl
\begin{thebibliography}{00}

\bibitem{QFTbook} M.~E.~Peskin and D.~V.~Schroeder, {\it An introduction to quantum field theory}, Perseus Books, Reading, Massachusetts 1995.

\bibitem{Kakubook} M. Kaku, {\it Quantum field theory}, Oxford University Press, New York 1993.

\bibitem{Wigner:1939cj}
  E.~P.~Wigner,
  Annals Math.\  {\bf 40} (1939) 149
   [Nucl.\ Phys.\ Proc.\ Suppl.\  {\bf 6} (1989) 9].
 
 \bibitem{BirrellDavies} N.~D.~Birrell and P.~C.~W.~Davies, {\it Quantum fields in curved space}, Cambridge University Press, Cambridge 1982.
  
\bibitem{Davies:1974th}
  P.~C.~W.~Davies,
  J.\ Phys.\ A {\bf 8} (1975) 609.
  
\bibitem{Unruh:1976db}
  W.~G.~Unruh,
  Phys.\ Rev.\ D {\bf 14} (1976) 870.
  
\bibitem{Hawking:1974sw}
  S.~W.~Hawking,
  Commun.\ Math.\ Phys.\  {\bf 43} (1975) 199,
  [Erratum: Commun.\ Math.\ Phys.\  {\bf 46} (1976) 206].
  
\bibitem{Hollands:2014eia}
  S.~Hollands and R.~M.~Wald,
  Phys.\ Rept.\  {\bf 574} (2015) 1
  [arXiv:1401.2026 [gr-qc]].
  
\bibitem{Fredenhagen:2014lda}
  K.~Fredenhagen and K.~Rejzner,
  J.\ Math.\ Phys.\  {\bf 57} (2016)
  031101
  [arXiv:1412.5125 [math-ph]].

\bibitem{Wohlfarth:2012ez}
  M.~N.~R.~Wohlfarth and C.~Pfeifer,
  Phys.\ Rev.\ D {\bf 87} (2013) 
  024031
  [arXiv:1210.4108 [gr-qc]].

\bibitem{Yanobook} K. Yano and S. Ishihara, {\it Tangent and cotangent bundles}, Marcel Dekker, New York 1973.

\bibitem{bookdiffeq} S.~Lang, {\it Differential manifolds}, Addison-Wesley, Reading, Massachusetts 1972.

\bibitem{Herzog:1995zz}
  T.~J.~Herzog, P.~G.~Kwiat, H.~Weinfurter and A.~Zeilinger,
  Phys.\ Rev.\ Lett.\  {\bf 75} (1995) 3034.

\bibitem{Elitzur:1993xh}
  A.~C.~Elitzur and L.~Vaidman,
  Found.\ Phys.\  {\bf 23} (1993) 987
  [hep-th/9305002].

\bibitem{Walborn:2002zza}
  S.~P.~Walborn, M.~O.~Terra Cunha, S.~Padua and C.~H.~Monken,
  Phys.\ Rev.\ A {\bf 65} (2002) 033818.
  
\bibitem{Bassi:2017szd}
  A.~Bassi, A.~Gro§ardt and H.~Ulbricht,
  Class.\ Quant.\ Grav.\  {\bf 34} (2017) no.19,  193002
  [arXiv:1706.05677 [quant-ph]].
  
\bibitem{BucataruMiron} I.~Bucataru and R.~Miron, {\it Finsler-Lagrange geometry, applications to dynamical systems}, Editura Academiei Romane, Bucuresti 2007. 

\end{thebibliography}
